\documentclass{svmult}
\usepackage{amsmath,amssymb}
\usepackage{graphicx}        % standard package for included graphics
\mainmatter  % start of an individual contribution
% ...
\begin{document}

\title*{Are Copying and Innovation Enough?\thanks{Contribution to
the proceedings of ECMI08, based on a talk given by A.D.K.Plato
as part of the minisymposium on Mathematics and Social
Networks. Preprint number \texttt{Imperial/TP/08/TSE/1}
\texttt{arXiv:0809.2568v1}}}

% Shortened title if needed
%\titlerunning{Copying and Innovation}

\author{T.S.Evans\inst{1,2}, A.D.K. Plato\inst{2} and T.You\inst{1}}
\institute{Theoretical Physics,
 Imperial College London,
 SW7 2AZ,  U.K.
\and Inst.\ for Mathematical Sciences, Imperial College London,
SW7 2PG, U.K.}

\maketitle
\begin{abstract}
% A very short abstract written in the third person
Exact analytic solutions and various numerical results for the
rewiring of bipartite networks are discussed. An interpretation
in terms of copying and innovation processes make this relevant
in a wide variety of physical contexts.  These include Urn
models and  Voter models, and our results are also relevant to
some studies of Cultural Transmission, the Minority Game and
some models of ecology.
\end{abstract}

% *******************************************************
\subsubsection*{Introduction}

There are many situations where an `individual' chooses only
one of many `artifacts' but where their choice depends in part
on the current choices of the community. Names for new babies
and registration rates of pedigree dogs often reflect current
popular choices \cite{HB03,HBH04}. The allele for a particular
gene carried (`chosen') by an individual reflects current gene
frequencies \cite{E04}. In Urn models the probabilities
controlling the urn chosen by a ball can reflect earlier
choices \cite{GL02}. In all cases copying the state of a
neighbour, as defined by a network of the individuals, is a
common process because it can be implemented without any global
information \cite{ES05}. At the other extreme, an individual
might might pick an artifact at random.

\subsubsection*{The Basic Model}

\begin{figure}[hbt]
  \sidecaption
  %\centering
%  \includegraphics[width=6cm]{copymodel3ind.eps}
  \includegraphics[width=0.5\linewidth]{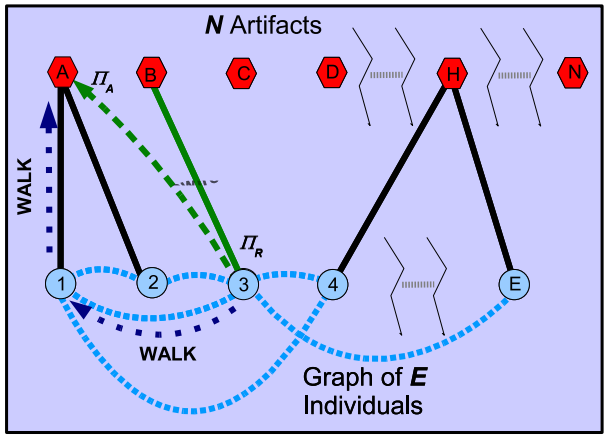}
\caption{The bipartite network of $E$ individual vertices,
each connected by a single edge (solid lines) to any one
of $N$ artifacts. The dashed lines below the individuals are a
social network. In the event shown individual \textbf{3} updates their choice,
making \textbf{B} the departure artifact.
They do this by copying the choice of a friend, friend of a friend, etc.,
found by making a random walk
on the social network.  Here this produces \textbf{A} as the arrival artifact so
edge \textbf{3B} is rewired to become edge \textbf{3A}.}
 \label{fCopyModel3ind}
\end{figure}

We first consider a non-growing bipartite network in which $E$
`individual' vertices are each attached by a single edge to one
of $N$ `artifact' vertices. At each time step we choose to
rewire the artifact end of one edge, the \emph{departure}
artifact chosen with probability $\Pi_R$. This is attached to
an \emph{arrival} artifact chosen with probability $\Pi_A$.
Only after both choices are made is the graph rewired as shown
in Fig. \ref{fCopyModel3ind}.
The degree distribution of the artifacts when averaged over
many runs of this model, $n(k,t)$, satisfies the following
equation:-
\begin{eqnarray}
n(k,t+1) &=& n(k,t) +
   n(k+1,t) \Pi_R(k+1,t) \left( 1- \Pi_A(k+1,t) \right)
 \nonumber \\ {}
 &&
     - n(k,t)   \Pi_R(k,t)   \left( 1- \Pi_A(k,t)   \right)
     - n(k,t)   \Pi_A(k,t)   \left( 1- \Pi_R(k,t)   \right)
 \nonumber \\ {}
 &&
     + n(k-1,t) \Pi_A(k-1,t) \left( 1- \Pi_R(k-1,t) \right)
   ,
   \;\;\; (E \geq k \geq 0)  \, ,
   \label{neqngen}
\end{eqnarray}
where $n(k)=\Pi_R(k)=\Pi_A(k)=0$ for $k=-1, (E+1)$.  If $\Pi_R$
or $\Pi_A$ have terms proportional to $k^\beta$ then this
equation is exact only when $\beta=0$ or $1$ \cite{EP07}. We
will use the most general $\Pi_R$ and $\Pi_A$ for which
(\ref{neqngen}) is exact, namely
\begin{equation}
 \Pi_R = \frac{k}{E}, \qquad
 \Pi_A = p_r\frac{1}{N} + p_p\frac{k}{E},
   \qquad p_p+p_r=1 \qquad (E \geq k \geq 0) \; .
 \label{PiRPiAsimple}
\end{equation}
This is equivalent to using a complete graph with self loops
for the social network at this stage but these preferential
attachment forms emerge naturally when using a random walk on a
general network \cite{ES05}. This choice for $\Pi_A$ has two
other special properties: one involves the scaling properties
\cite{EP07} and the second is that these exact equations can be
solved analytically \cite{Evans07,EP07,EP07eccs07,Evans07pisa}.
The generating function $G(z,t) = \sum_k z^k n(k,t)$ is
decomposed into eigenmodes $G^{(m)}(z)$ through $G(z,t) =
\sum_{m=0}^{E} c_m (\lambda_m)^t G^{(m)}(z)$. From
(\ref{neqngen}) we find a second order linear differential
equation for each of the eigenmodes with solution \cite{EP07}
\begin{eqnarray}
 G^{(m)}(z) &=&  (1-z)^m {}_2F_1(a+m,-E+m;1-E-a(N-1) ;z) \,
 , \;\;\; a=\frac{p_r}{p_p}\frac{E'}{N} \, ,
 \nonumber \\ {}
 \lambda_{m} &=& 1 -m(m-1) \frac{p_p}{E E'} - m \frac{p_r}{E},
 \qquad 0 \leq m \leq E \; ,
 \label{Gmresult}
 \label{evalues}
\end{eqnarray}
where $E'=E$. These solutions are well known in theoretical
population genetics as those of the Moran model \cite{E04} and
one may map the bipartite model directly onto a simple model of
the genetics of a haploid population \cite{EP07}.

The equilibrium result for the degree distribution
\cite{Evans07,EP07} is proportional to $\frac{\Gamma\left( k +
a \right) }{\Gamma \left(k+1\right) } \frac{\Gamma\left(
E+a(N-1)-1 - k\right) }{\Gamma \left(E  +1  -  k  \right) }$.
This has three typical regions. We have a condensate, where
most of the edges are attached to one artifact $p(k=E) \sim
O(N^0)$, for $p_r \ll (E+1-\langle k \rangle)^{-1}$ . On the
other hand when $p_r\gg (1+\langle k \rangle)^{-1}$ we get a
peak at small $k$ with an exponential fall off, a distribution
which becomes an exact binomial at $p_r=1$.  In between we get
a power law with an exponential cutoff, $p(k) \propto
(k)^{-\gamma} \exp\{-\zeta k\}$ where $\gamma \approx
(1-\frac{p_r}{p_p}\langle k \rangle)$ and $\zeta \approx
-\ln(1-p_r)$.  For many parameter values the power $\gamma$
will be indistinguishable from one and this is a characteristic
signal of an underlying copying mechanism seen in a diverse
range of situations (e.g.\ see \cite{ATBK04,LJ06b}).
\begin{figure}[htb!]
  \sidecaption
  \includegraphics[width=0.5\linewidth]{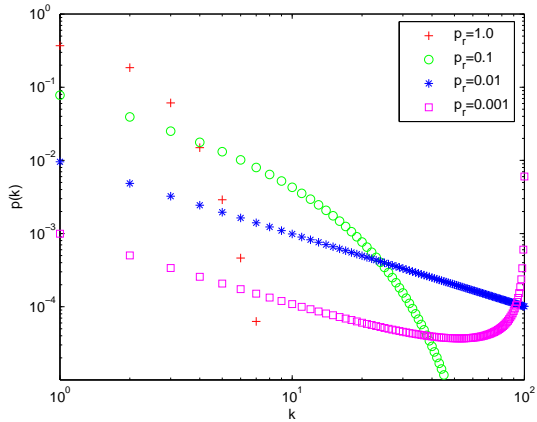}
      \caption{The equilibrium degree probability distribution
function $p(k)=n(k)/N$ for $N=E=100$.  Shown are (from top to bottom at low $k$)
$p_r=1$ (red crosses), $10/E$ (green circles), $1/E$ (blue stars) and $0.1/E$
(magenta squares).}\label{feqDD}
 \end{figure}

One of the best ways to study the evolution of the degree
distribution \cite{EP07,EP07eccs07} is through the
\emph{Homogeneity Measures}, $F_n$. This is the probability
that $n$ distinct edges chosen at random are connected to same
artifact, and is given by $F_n(t) :=
(\Gamma(E+1-n)/\Gamma(E+1)) (d^nG(z,t)/dz^n)_{z=1}$. Further,
each $F_n$ depends only on the modes numbered $0$ to $n$ so
they provide a practical way to fix the constants $c_n$ in the
mode expansion.  Since $F_0=E$ and $F_1=1$, we find $c_0=1$ and
$c_1=0$ while equilibration occurs on a time scale of $\tau_2=
-1/\ln(\lambda_2)$.
\begin{figure}
\sidecaption
 %\centering
%  \includegraphics[width=6cm]{Fnnina100pr0_01.eps}
  \includegraphics[width=0.5\linewidth]{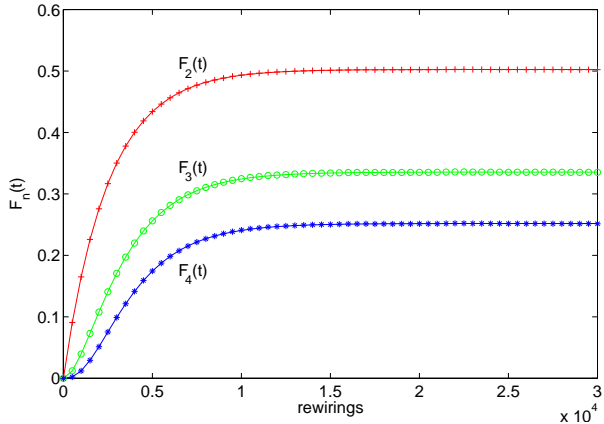}
      \caption{Plots of various $F_n(t)$ for $E=N=100$, $p_r=0.01$. The
points are averages over $10^5$ runs while the lines are the exact
theoretical results. From top to bottom we have: $F_2(t)$
(crosses), $F_3(t)$ (circles), $F_4(t)$ (stars).} \label{fnVarious}
\end{figure}

\subsubsection*{Communities}

Our first generalisation of the basic model is to consider two
distinct communities of individuals, say $E_x$ ($E_y$) of type
X (Y). The individuals of type X can now copy the choices made
by their own community X with probability $p_{pxx}$, but a
different rate is used when an X copies the choice made by
somebody in community Y, $p_{pxy}$. An X individual will then
innovate with probability $(1-p_{pxx}-p_{pxy})$. Another two
independent copying probabilities can be set for the Y
community. At each time step we choose to update the choice of
a member of community X (Y) community with probability $p_x$
($1-p_x$). Complete solutions are not available but one can
find exact solutions for the lowest order Homogeneity measures
and eigenvalues using similar techniques to those discussed
above. The unilluminating details are given in
\cite{EP07eccs07}.

\subsubsection*{Complex Social Networks}

An obvious generalisation is to use a complex network as the
Individual's social network \cite{EP07eccs07}. When copying,
done with probability $p_p$, an individual does a random walk
on the social network to choose another individual and finally
to copy their choice of artifact, as shown in
Fig.~\ref{fCopyModel3ind}. The random walk is an entirely local
process, no global knowledge of the social network is needed,
so it is likely to be a good approximation of many processes
found in the real world.  It also produces an attachment
probability which is, to a good approximation, proportional to
the degree distribution \cite{ES05}.  The alternative process
of innovation, followed with probability $p_r$, involves global
knowledge through its normalisation $N$ in
(\ref{PiRPiAsimple}). However when $N \gg E$ this can represent
innovation of new artifacts as it is likely that the arrival
artifact has never been chosen before. However this process
could also be a first approximation for other unknown processes
used for artifact choice.

\begin{figure}[hbt!]
 \sidecaption
 \centering
 \includegraphics[width=0.5\linewidth]{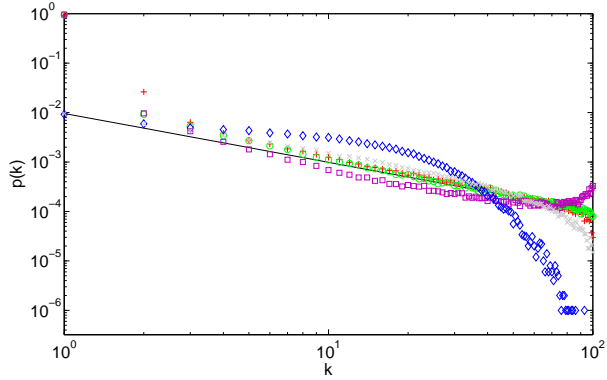}
\caption{The degree distributions
$p(k)$ averaged over $10^4$ runs for different social networks
of average degree of 4: Erd\H{o}s-R\'{e}yni (red
pluses), Exponential (random with $p(k) \propto \exp(-\zeta k)$,
green circles), Scale Free (random with $p(k) \propto k^{-3}$,
purple squares), periodic lattices of two (grey crosses) and one
(blue diamonds) dimension.
The line is the analytic result where the social network is a complete graph with self loops.
$N=E=100$, $p_r=1/E$.}
 \label{fsocialnet}
\end{figure}
Results shown in Fig.\ref{fsocialnet} show that the existence
of hubs in the Scale Free social network enhances the
condensate while large distances in the social networks, as
with the lattices, suppress the condensate.

\begin{figure}[hbt!]
 \sidecaption
 \centering
\includegraphics[width=0.5\linewidth]{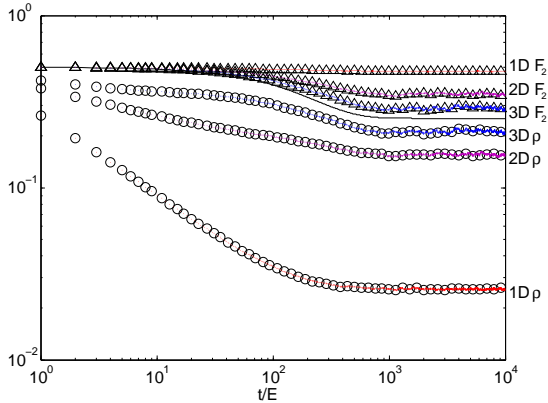}
\caption{Inhomogeneity measures for various lattices against $t/E$.
The black solid line represents the analytic result $(1-F_2(t))$ for
$N=2$, $p_r=1/E$ and $E=729$. Numerical results for $(1-F_2(t))$
(triangles) and for the average probability that a neighbour has a different artifact,
$rho(t)$ (circles) shown for social networks which are lattices of different dimensions. Averaged over $1000$ runs.}
 \label{flatcomp}
\end{figure}

An interesting example is the case of $N=2$ which is a Voter
Model \cite{Liggett99} with noise (innovation $p_r\neq 0$)
added. One can then compare the probability that a neighbour
has a different artifact (the interface density) $\rho(t)$, a
local measure of the inhomogeneity, with our global measure
$(1-F_2(t))$. These coincide when the social network is a
complete graph. However as we move from 3D to 1D lattices,
keeping $N$, $E$ and $p_r$ constant, we see from
Fig.~\ref{flatcomp} that both these local and global measures
move away from the result for the complete graph but in
opposite directions \cite{EP07eccs07}.

\subsubsection*{Different Update Methods}

Another way we can change the model is to change the nature of
the update.  Suppose we first select the edge to be rewired and
immediately remove it. Then, based on this network of
$E'=(E-1)$ edges, we choose the arrival artifact with
probability $\Pi_A=(p_r/N)+ (1-p_r)k/E'$. The original master
equation (\ref{neqngen}) is still valid and exact. Moreover it
can still be solved exactly giving exactly the same form as
before, (\ref{Gmresult}), but with $E'=(E-1)$ not $E$. This
gives very small differences of order $O(E^{-1})$ when compared
to the original simultaneous update used initially.

Instead we will consider the simultaneous rewiring of $X$ edges
in our bipartite graph at each step. We will choose the
individuals, whose edges define the departure artifacts, in one
of two ways: either sequentially or at random. The arrival
artifacts will be chosen as before using $\Pi_A$ of
(\ref{PiRPiAsimple}).

\begin{figure}[htb]
\centering
\includegraphics[width=7.0cm]{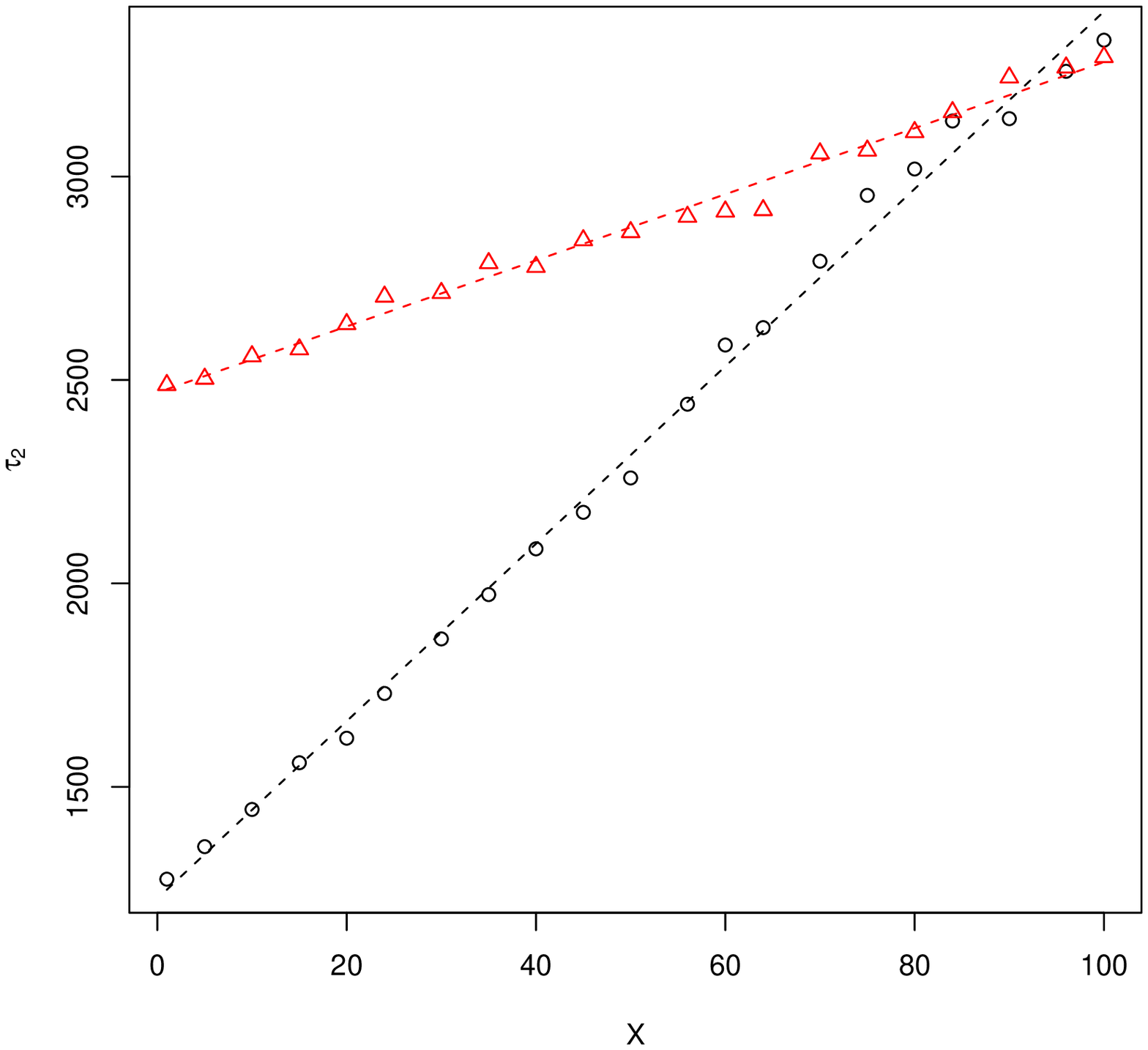}
\includegraphics[width=7.0cm]{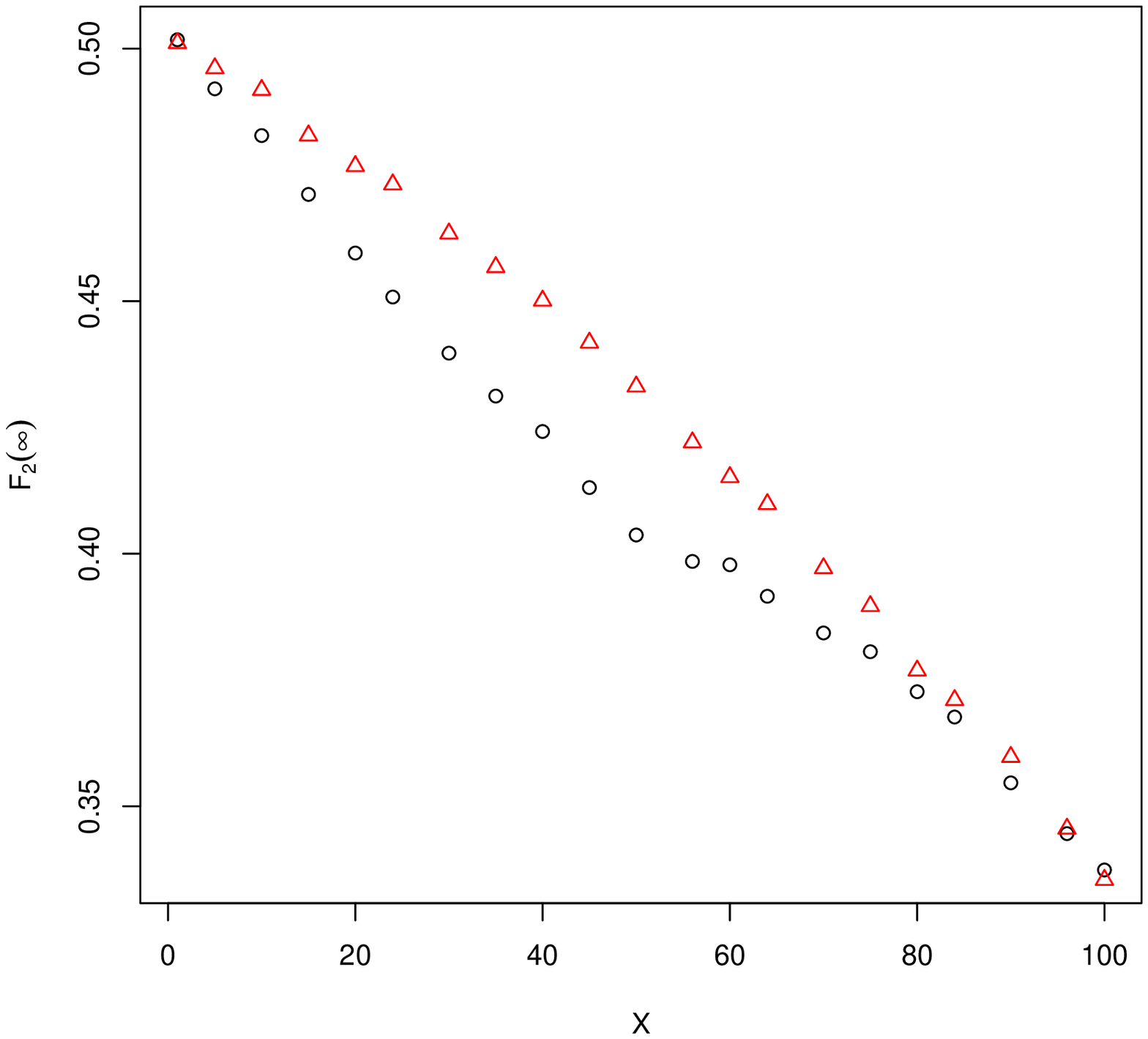}
\caption{$\tau_2= - 1/\ln(\lambda_2)$ (left) and $F_2(\infty)$ (right) obtained by
fitting $A+B(\lambda_2)^t$ to the data for $F_2(t)$. For
sequential ($m=4$ black circles, lower lines) and random ($m=6$ red triangles, upper lines)
updates of $X$ individuals at a time. $N=E=100$, $p_r = 1/E
=0.01$ and averaged over $10^4$ runs. The dashed lines represent
the best linear fit with $\tau_2 \approx 1230 (20) + 21.8(3)X$ for
$m=4$ and $\tau_2 \approx 2470 (10) + 8.1(2)X$ for $m=6$.
Theoretical values are $\tau_2 \approx 2512.1$ and $F_2(\infty) \approx 0.50251$ for $X=1$ random update
and $\tau_2 \approx 3316.6$ and $F_2(\infty) \approx 0.33669$ for $X=100$ either update.}
\label{f2tau2pr01}
\end{figure}

The opposite extreme from the single edge rewiring case we
started with ($X=1$) is the one where all the edges are rewired
at the same time, $X=E$.  This is the model used in
\cite{HB03,HBH04,BLHH07} to model various data sets on cultural
transmission. It is also the classic Fisher-Wright model of
population genetics \cite{E04}. From this each homogeneity
measure $F_n$ and the $n$-th eigenvector $\lambda_n$ may be
calculated in terms of lower order results $F_m$ ($m<n$). Non
trivial information again comes first from $F_2(t) =
F_2(\infty) + (\lambda_2)^t \left(F_2(0)-F_2(\infty)\right)$
where
\begin{eqnarray}
 F_2(\infty) &=& \frac{p_p^2 + (1-p_p^2)\langle k \rangle}{p_p^2 + (1-p_p^2)E}
 \, , \qquad
 \lambda_2 = \frac{p_p^2(E-1)}{E} \, .
\end{eqnarray}
Comparing with the results for $X=1$ we see that there are
large differences in the equilibrium solution and in the rate
at which this is approached (measured in terms of number of the
rewirings made). For intermediate values of $X$ we have not
obtained any analytical results so for these numerical
simulations are needed, as shown in Fig.\ref{f2tau2pr01}.

\newpage

% *****************************************************************

%%%%%%%%%%%%%%%%%%%%%%%%%%%%%%%%%%%%%%%%%%%%%%%%%%%%%%%%%%%%%%%%%%%%%%
%\printindex

\newpage
\section*{Supplementary Material}

This following material is not part of the published paper.

\begin{figure}[hbt]
\centering
\includegraphics[width=0.5\linewidth]{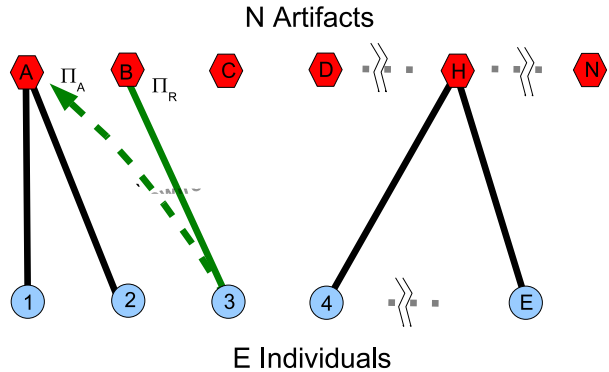}
\caption{Each of $E$ `individual' vertices is connected by a
single edge to one of $N$ `artifact' vertices. In the simplest
model the artifact end of one edge is rewired at each time step.
The edge to be rewired is chosen with probability $\Pi_R$ (the
edge from individual $3$ to the departure artifact D). At the same
time the arrival artifact is chosen with probability $\Pi_A$ (here
labelled A). The rewiring is performed only after both choices
have been made.}
 \label{fCopyModel3}
\end{figure}

Fig.\ref{fCopyModel3} shows the basic model --- simultaneous
rewiring of the artifact end of a single edge.

\begin{figure}[hbt] {\centerline{
 \includegraphics[width=0.5\linewidth]{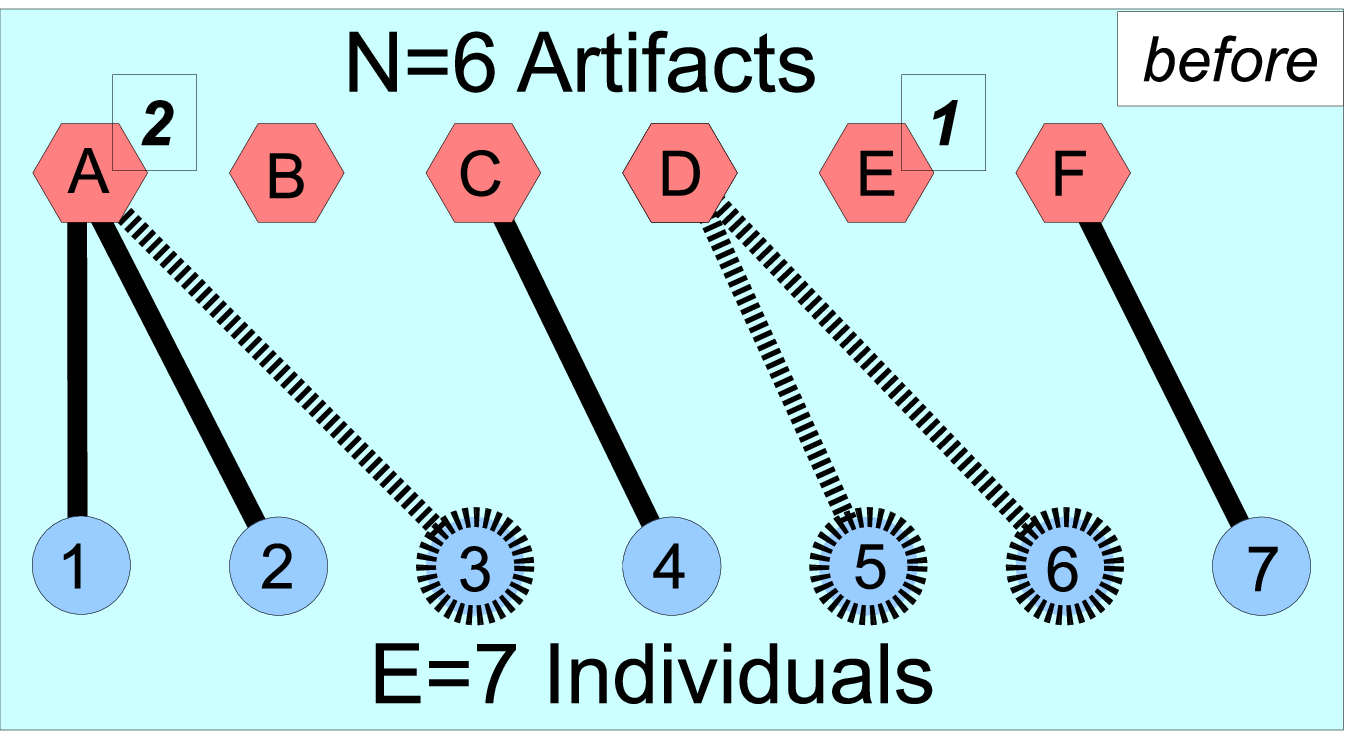}
 \includegraphics[width=0.5\linewidth]{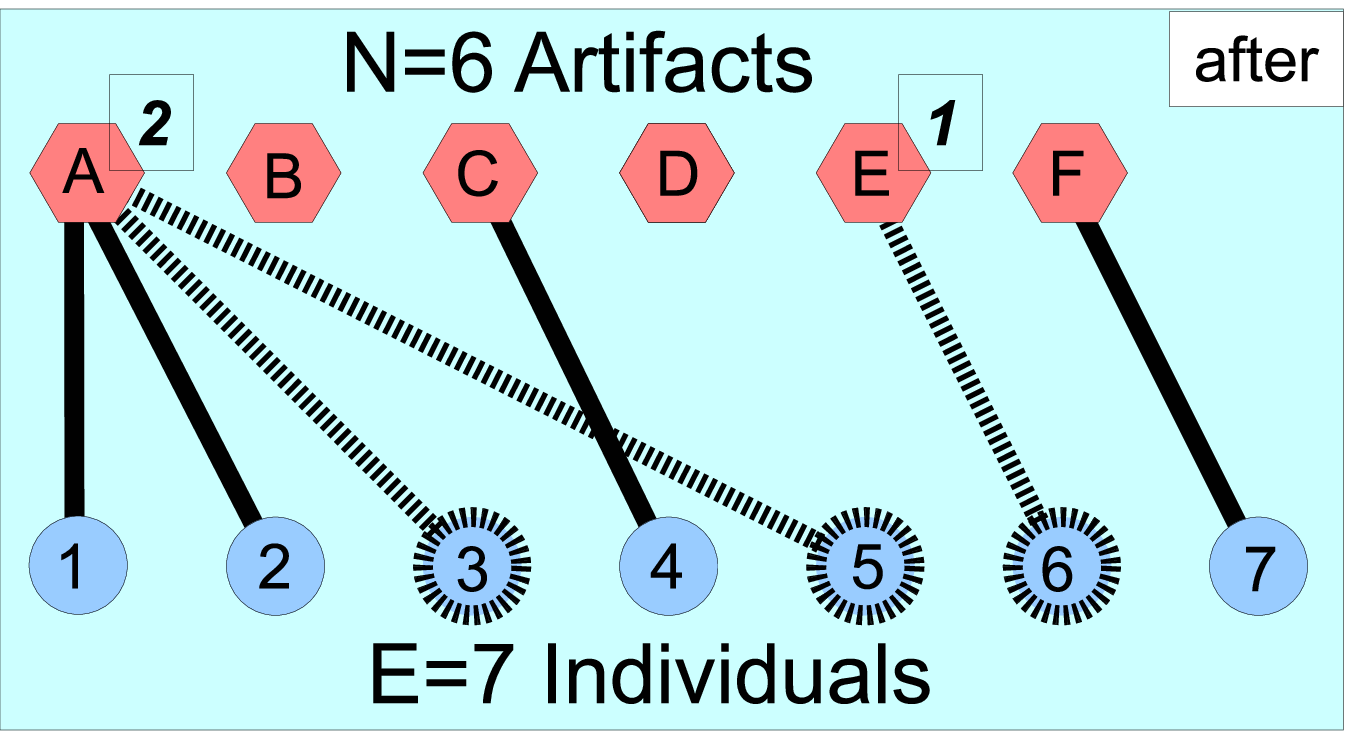}} }
\caption{An example of the rewiring of the bipartite graph.
Here the choice made by each of the $E=7$ `individual' vertices
is represented by an edge connected to one of the $N=6$
`artifact' vertices. At each time step $X$ individuals decide
to change their choice.   Here $X=3$ and the chosen individuals
(\textbf{3},\textbf{5}, and \textbf{6}) and their edges are
indicated by dashed lines in the left hand panel. The new
artifacts for the $X$ individuals are chosen with probability
$\Pi_A$. Here \textbf{A} is chosen twice and \textbf{E} once and
the result is shown on the right.}
 \label{fCopyModelX}
\end{figure}

The evolution equation when $X$ edges are rewired
simultaneously, as shown in Fig.\ref{fCopyModelX}, is
\begin{eqnarray}
  G(z,t+E)
  &=& \sum_{k'=0}^E [1+(z-1)\Pi_\mathrm{A}(k')]^E n(k',t)  \, .
 \label{Gpevol}
\end{eqnarray}

\begin{figure}
\centering
\includegraphics[width=10.0cm]{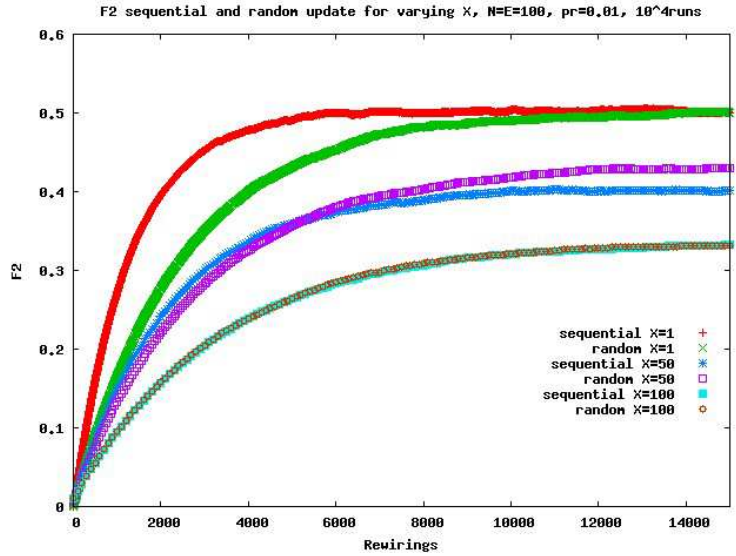}
\caption{$F_2$ for sequential and random
updates of $X$ individuals at a time.  $N=E=100$, $p_r = 1/E$
and averaged over $10^4$ runs. }
 \label{fktp}
\end{figure}
For the values used in Fig.\ \ref{f2tau2pr01} and \ref{fktp},
we would predict $F_2(X=100;\infty)=0.3367 \approx (1/3)
+O(1/E)$ while $F_2(\mathrm{Rand},X=1;\infty)=0.5025 \approx
(1/2) +O(1/E)$. These clearly match the numerical results shown
in Fig.\ \ref{f2tau2pr01} and  \ref{fktp}.

\subsubsection*{Summary}

We have shown how simple models of bipartite network rewiring
can be solved exactly.  The preferential attachment can be seen
as emerging from simple copying using local information only on
the social network. On the other hand the large $N$ limit shows
the random attachment process may be thought of as innovation.
Many other models can be mapped to this simple network model
--- see the review in \cite{EP07}.  Thus copying and innovation may be
enough to explain the results seen in many other contexts. such
as the Minority game \cite{ATBK04} and in models of evolution
\cite{LJ06b}.

\end{document}